\documentclass[12pt]{iopart}
\usepackage{graphicx}
\usepackage{iopams}
\newcommand{\be}{\begin{equation}}
\newcommand{\ee}{\end{equation}}

\newcommand{\mnras}{Mon. Not. Roy. Astr. Soc. }

\begin{document}

\title{Dark Matter: The evidence from astronomy, astrophysics and cosmology}

\author{Matts Roos}
\address{Department of Physics, FI-00014 University of Helsinki, Finland}
\ead{matts.roos@helsinki.fi}
\date{\today}

\begin{abstract}
Dark matter has been introduced to explain many independent gravitational effects at different astronomical scales, in galaxies, groups of galaxies, clusters, superclusters and even across the full horizon. This review describes the accumulated astronomical, astrophysical, and cosmological evidence for dark matter. It is written at a non-specialist level and intended for an audience with little or only partial knowledge of astrophysics or cosmology.
\end{abstract}
 \maketitle

\section*{Contents}
1. Introduction\\
2. Stars near the Galactic disk\\
3. Virially bound systems\\
4. Rotation curves of spiral galaxies\\
5. Small galaxy groups emitting X-rays\\
6. Mass to luminosity ratios\\
7. Mass autocorrelation functions\\
8. Strong and weak lensing\\
9. Cosmic Microwave Background\\
10. Baryonic acoustic oscillations\\
11. Galaxy formation in purely baryonic matter\\
12. Large Scale Structures simulated\\
13. Dark matter from overall fits\\
14. Merging galaxy clusters\\
15. Comments and conclusions

\section{Introduction}

According to the standard ``Cold Dark Matter'' (CDM) paradigm,
dark matter originated from quantum fluctuations in the almost uniform, early Universe during a very short period of cosmic inflation, and emerged with negligible thermal velocities and a Gaussian and scale-free distribution of density fluctuations. It
is one of the most important tasks of our time to establish whether this paradigm is supported by observations.

In spite of the importance of this task, there is no review of the accumulated astronomical, astrophysical, and cosmological observational testimony on the CDM paradigm. The literature summing such evidence usually refers only to the rotation curves of spiral galaxies, to the observations of strongly lensing galaxies and clusters, and to the determination of parameters in fits combining the Cosmic Microwave Background (CMB) with some other constraints. Quite recently the ``Bullet cluster'' gained fame due to the combination of lensing observations of its gravitational field and X-ray observations of the heated Intra-Cluster Medium (ICM). But there exists much more evidence.

The purpose of this review is to summarize all the gravitational evidence of dark matter without entering into speculation on its true nature. Whether the correct explanation implies conventional matter, unconventional particles, or modifications to gravitational theory, all the observed gravitational effects described below have to be explained. Thus the use of the term ``dark matter'' is generic and does not imply that we are taking sides.

Beginning from the first startling discoveries of Jan Hendrik Oort in 1932 \cite{Oort1} of missing matter in the Galactic disk which has not been confirmed by modern observations (Sec. 2), and that of Fritz Zwicky in 1933 \cite{Zwicky} of missing matter in the Coma cluster (Sec. 3), much later understood to be ``dark matter'', we describe the kinematics of virially bound systems (Sec. 3) and rotating spiral galaxies (Sec. 4). Different testimonies of missing mass in groups and clusters are offered by the comparison of visible light and X-rays (Sec. 5). The mass to luminosity ratios of objects at all scales can be compared to what would be required to close the Universe (Sec. 6), and mass autocorrelation functions relate galaxy masses to dark halo masses (Sec. 7). An increasingly important method to determine the weights of galaxies, clusters and gravitational fields at large, independently of electromagnetic radiation, is lensing, strong as well as weak (Sec. 8).

In radiation the most important tools are the temperature and polarization anisotropies in the Cosmic Microwave Background (Sec. 9), which give information on the mean density of both dark matter and baryonic matter as well as on the geometry of the Universe. The large scale structures of matter exhibit similar fluctuations with similar information in the Baryonic Acoustic Oscillations (BAO) (Sec. 10). The amplitude of the temperature variations in the CMB prove, that galaxies could not have formed in a purely baryonic Universe (Sec. 11). Simulations of large scale structures also show that dark matter must be present (Sec. 12). The best quantitative estimates of the density of dark matter come from overall parametric fits (to various cosmological models) of CMB data, BAO data, and redshifts of supernovae of type Ia (SNe\,Ia) (Sec. 13). A particularly impressive testimony comes from combining lensing information on pairs of colliding clusters with X-ray maps (Sec. 14). We conclude (Sec. 15) with a few general comments on the nature of dark matter.

\section{Stars near the Galactic disk}
In 1932 the Dutch astronomer Jan Hendrik Oort analyzed the vertical motions of all known stars near the Galactic plane and used these data to calculate the acceleration of matter \cite{Oort1}. This amounts to treating the stars as members of a "star atmosphere", a statistical ensemble in which the density of stars and their velocity dispersion defines a "temperature" from which one obtains the gravitational potential.  This is analogous to how one obtains the gravitational potential of the Earth from a study of the atmosphere. The result contradicted grossly the expectations: the potential provided by the known stars was not sufficient to keep the stars bound to the Galactic disk, the Galaxy should rapidly be losing stars. Since the Galaxy appeared to be stable there had to be some missing matter near the Galactic plane, Oort thought, exerting gravitational attraction. This used to be counted as the first indication for the possible presence of dark matter in our Galaxy.

The possibility that this missing matter would be non-baryonic could not even be thought of at that time. Note that the first neutral baryon, the neutron, was discovered by James Chadwick in the same year, in 1932.

However, it is nowadays considered, that this does not prove the existence of dark matter in the disk. The potential in which the stars are moving is not only due to the disk, but rather to the totality of matter in the Galaxy which is dominated by the Galactic halo. The advent of much more precise data in 1998 led Holmberg and Flynn  \cite{Holmberg} to conclude that no dark matter was present in the disk.

Oort determined the mass of the Galaxy to be $10^{11}~\rm M_{sun}$, and thought that the nonluminous component was mainly gas. Still in 1969 he thought that intergalactic gas  made up a large fraction of the mass of the universe \cite{Oort2}. The general recognition of the missing matter as a possibly new type of non-baryonic dark matter dates to the early eighties.

\section{Virially bound systems}
Stars move in galaxies and galaxies in clusters along their orbits; the orbital velocities are balanced by the total gravity of the system, similar to the orbital velocities of planets moving around the Sun in its gravitational field. In the simplest dynamical framework one treats clusters of galaxies as statistically steady, spherical, self-gravitating systems of $N$ objects of average mass $m$ and average orbital velocity $v$. The total kinetic energy $E$ of such a system is then
\begin{equation}
E={1\over 2}Nmv^2\ .
\end{equation}
If the average separation is $r$, the potential energy of  $N(N-1)/2$ pairings is
\begin{equation}
U=-{1\over 2}N(N-1){Gm^2\over r}\ .
\end{equation}
The virial theorem states that for such a system
\begin{equation}
E=-U/2\ .\label{vir}
\end{equation}
The total dynamic mass $M$ can then be estimated from $v$, and $r$ from the cluster volume
\begin{equation}
M=Nm={2rv^2\over G}\ .
\end{equation}

An empirical formula often used in simulations of the density distribution of dark matter halos in clusters is
\begin{equation}
\rho_{DM}(r)=\frac{\rho_0}{(r/r_s)^{\alpha}(1+r/r_s)^{3-\alpha}}\ ,\label{halo}
\end{equation}
where $\rho_0$ is a normalization constant and $0\leq\alpha\leq 3/2$. Standard choices are $\alpha=1$ for the Navarro-Frenk-White profile \cite{NFW} and the $\alpha=3/2$ profile of Moore \& al. \cite{Moore}, both cusped at $r=0$.

However, Gentile \& al. \cite{Gentile} have shown that cusped profiles are in clear conflict with data on spiral galaxies. Central densities are rather flat, scaling approximately as $\rho_0\propto r^{-2/3}_{luminous}$. The best-fit disk + NFW halo mass model fits the rotation curves poorly, it implies an implausibly low stellar mass-to-light ratio and an unphysically high halo mass.

\subsection{The Coma cluster}
Historically, the second possible indication of dark matter, the first time in an object at a cosmological distance, was found by Fritz Zwicky in 1933 \cite{Zwicky}. While measuring radial velocities of member galaxies in the Coma cluster (that contains some $1000$ galaxies), and the cluster radius from the volume they occupy, Zwicky was the first to use the virial theorem to infer the existence of unseen matter. He was able to infer the average mass of galaxies within the cluster, and obtained a value about 160 times greater than expected from their luminosity (a value revised today), and proposed that most of the missing matter was dark.
   \begin{figure}[htbp]
\includegraphics[width=12cm]{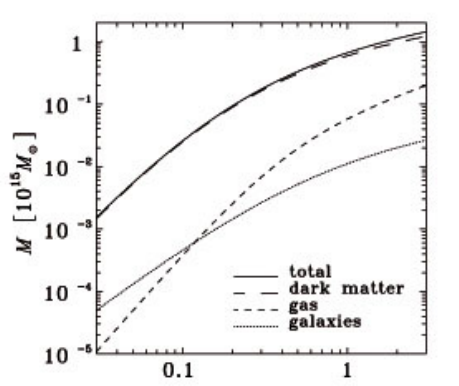}
\caption{Density profile of matter components enclosed within a given radius $r$ in the Coma cluster versus $r/r_{virial}$. From E. L. Lokas \& G. A. Mamon \cite{Lokas}.}
\end{figure}
His suggestion was not taken seriously at first by the astronomical community, that Zwicky felt as hostile and prejudicial. Clearly, there was no candidate for the dark matter because gas radiating X-rays and dust radiating in the infrared could not yet be observed, and non-baryonic matter was unthinkable.  Only some forty years later when studies of motions of stars within galaxies also implied the presence of a large halo of unseen matter extending beyond the visible stars, dark matter became a serious possibility.

Zwicky found to his surprise that the orbital velocities were almost a factor of ten
larger than expected from the summed mass of all galaxies belonging
to the Coma cluster. He then concluded that in order to hold galaxies together the cluster must contain huge amounts of some non-luminous matter.
Since that time modern observations have revised our understanding of the composition of clusters. Luminous stars represent a very small fraction of a cluster mass; in addition there is a baryonic, hot intracluster medium (ICM) visible in the X-ray spectrum. Rich clusters typically have more mass in hot gas than in stars; in the largest virial systems like the Coma the composition is about 85\% dark matter, 14\% ICM, and only 1\% stars \cite{Lokas}.

In modern applications of the virial theorem one also needs to model and parametrize the radial distributions of the ICM and the dark matter densities. In the outskirts of galaxy clusters the virial radius roughly separates bound galaxies from galaxies which may either be infalling or unbound. The virial radius $r_{vir}$ is conventionally defined as the radius within which the mean density is 200 times the background density.

Matter accretion is in general quite well described within the approximation of the Spherical Collapse Model. According to this model, the velocity of the infall motion and the matter overdensity are related. Mass profile estimation is thus possible once the infall pattern of galaxies is known \cite{Cupani}.

In Fig. 1 the Coma profile is fitted \cite{Lokas} with Eq.~(\ref{halo}) with $\alpha=0$ which describes a centrally finite profile which is almost flat. The separation of different components in the core is not well done with Eq.~(\ref{halo}) because the Coma has a binary center like many other clusters \cite{Coma}.

The physical size of clusters is expected to depend on the mass, characterized by the mass concentration index $c_s\equiv r_{vir}/r_s$, roughly $c_s\propto M^{-0.1}$. This can be tested by strong lensing of a sequence of halos.
   \begin{figure}[htbp]
\includegraphics[width=12cm]{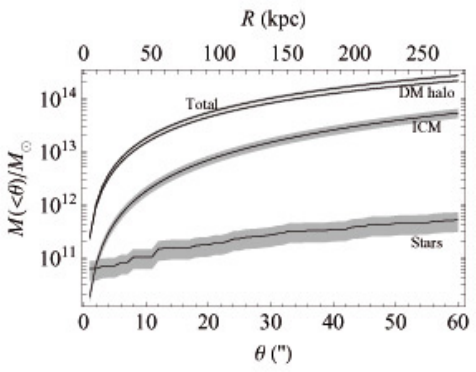}
\caption{Density profile of matter components in the cluster AC 114, enclosed within a given projected radius. From M. Sereno \& {\em al.} \cite{Sereno}.}
\end{figure}

\subsection{The AC 114 cluster}
Dark matter is usually dissected from baryons in lensing analyses by first fitting the lensing features to obtain a map of the total matter distribution and then subtracting the gas mass fraction as inferred from X-ray observations \cite{Bradac, Allen}.
The total mass map can then be obtained with parametric models in which the contribution from cluster-sized DM haloes can be considered together with the main galactic DM haloes \cite{Limousin}. Mass in stars and stellar remnants is estimated converting galaxy luminosity to mass assuming suitable stellar mass to light ratios.

One may go one step further by exploiting a parametric model which has three kinds of components: cluster-sized dark matter haloes, galaxy-sized (dark plus stellar) matter haloes, and a cluster-sized gas distribution \cite{Coma, Sereno}. The results of such an analysis of the dynamically active cluster AC 114 is shown in Fig. 2.

\subsection{The Local Group}
The Local Group is a very small virial system, dominated by two large galaxies, the M31 or Andromeda galaxy, and the Milky Way. The M31 exhibits blueshift, falling in towards us. Evidently the Galaxy and M31, as well as all the minor galaxies in the Local Group, form a bound system which is oscillating.

In this virial system the two large galaxies dominate the dynamics, so that it is not meaningful to define, as for the Coma, an average pairwise separation between galaxies, an average mass or an average orbital velocity. The total kinetic energy $E$ is still given by the sum over all the group members, and the potential energy $U$ by the sum over all the galaxy pairs, but here the pair formed by the M31 and the Milky Way dominates, and the pairings of the small members with each other are negligible.

An interesting recent claim is, that the mass estimate of the Local Group is also affected by the accelerated expansion, the ``dark energy''. A. D. Chernin \& al. \cite{Valtonen} have shown that the the potential energy $U$ is reduced in the force field of dark energy, so that the virial theorem for $N$ masses $m_i$ with baryocentric radius vectors ${\bf r}_i$ takes the form
\begin{equation}
E=-{1\over 2}U + U_2\ ,
\end{equation}
where $U$ is defined as in Eq.~(\ref{vir}), and
\begin{equation}
U_2=-\frac{4\pi\rho_v}{3}~\sum m_i{\bf r}_i^2
\end{equation}
is a correction which reduces the potential energy due to the
background dark energy density $\rho_v$. In the Local Group this correction to the mass appears to be quite substantial, of the order of $30\% -50\%$.

The dynamical mass of the local group comes out to be $3.2 - 3.7~ 10^{12}$ solar masses whereas the total visible mass of the Galaxy + M31 is only $2~10^{11}$
solar masses. Thus there is a large amount of dark matter.

   \begin{figure}[htbp]
\includegraphics[width=15cm]{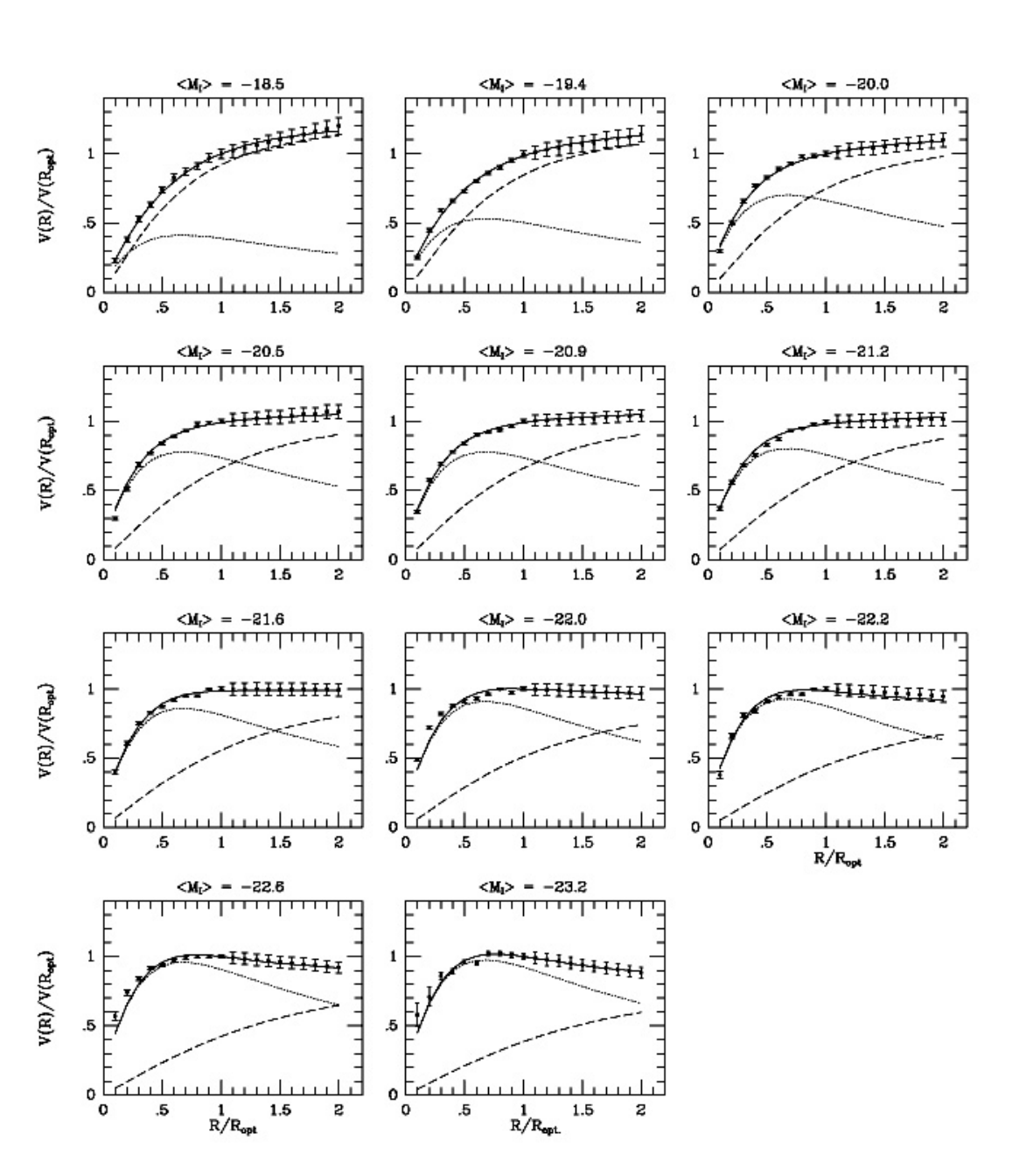}
\caption{Best disk-halo fits to the Universal Rotation Curve (dotted/dashed line: disc/halo). Each object is identified by the halo virial mass \cite{Salucci}.}
\end{figure}

\section{Rotation curves of spiral galaxies}
Spiral galaxies are stable gravitationally bound systems in which visible matter is composed of stars and interstellar gas. Most of the observable matter is in a relatively thin disc, where stars and gas rotate around the galactic center on
nearly circular orbits. If the circular velocity at
radius $r$ is $v$ in a galaxy with mass~$M(r)$ inside $r$, the condition for stability is that the centrifugal acceleration $v/r$ should equal the gravitational pull $GM(r)/r^2$, and the radial dependence of $v$ would then be expected to follow Kepler's law
\begin{equation}
v=\sqrt{\frac{GM(r)}{r}}.\label{disk}
\end{equation}
The surprising result from measurements of galaxy-rotation curves is that the velocity does not follow this $1/\sqrt{r}$ law, but stays constant after attaining a maximum at about 5~kpc. The most obvious solution to this is that the galaxies
are embedded in extensive, diffuse haloes of dark matter. If the enclosed
mass inside the radius r, M(r), is proportional to r
it follows that $v(r)\thickapprox$ constant.

The rotation curve of most galaxies can be fitted by the superposition of contributions from the stellar and gaseous disks, the luminous components, both modeled by thin exponential disks, sometimes
a bulge, and the dark halo, modeled by a quasi-isothermal sphere. The inner part is difficult to model because the density of stars is high, rendering observations of individual star velocities difficult. Thus the fits are not unique, the relative contributions of disk and dark matter is model-dependent, and it is not even sure whether galactic disks do contain dark matter. Typically  the dark matter is about half of the total mass.

In Fig. 3 we show the rotation curves fitted for eleven well-measured galaxies \cite{Salucci}. One notes, that at very small radii the dark halo component is indeed much smaller than the luminous disk component. At large radii, however, the need for a halo of dark matter is obvious. On galactic scales, the contribution of dark matter generally dominates the total mass. Note the contribution of the baryonic component, negligible for small masses but increasingly important in the larger structures.

Our Galaxy is complicated because of what appears to be a density dip at 9 kpc and a small dip at 3 kpc, as is seen in Fig. 4  \cite{Sofue}. To fit the measured rotation curve one needs at least three contributing components: a central bulge, the star disk + gas, and a dark matter  halo \cite{Sofue, Ullio, Weber}.  For small radii there is a choice of empirical rotation curves, and no dark matter component appears to be needed until radii beyond 15 kpc.

\begin{figure}[htbp]
   \includegraphics[width=15cm]{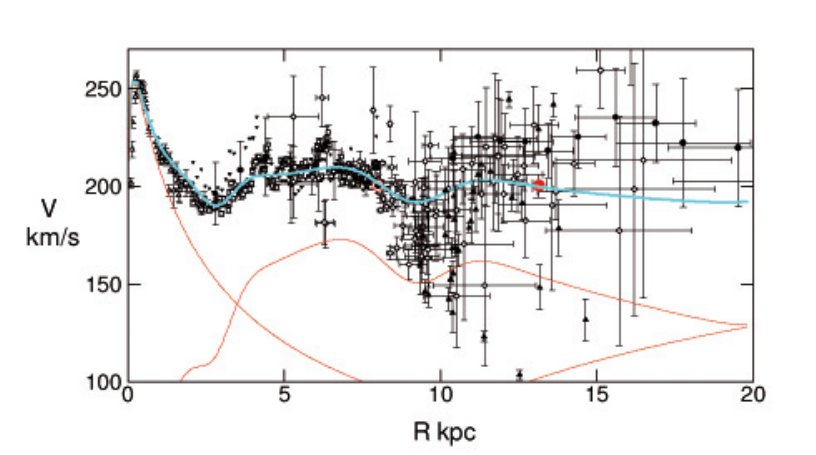}
   \caption{Decomposition of the rotation curve of the Milky Way into the components bulge, stellar disk + interstellar gas, dark matter halo (the red curves from left to right). From Y. Sofue {\em et al.} \cite{Sofue}.}
      \end{figure}

\newpage
\section{Small galaxy groups emitting X-rays}
There are examples of groups formed by a small number of galaxies
which are enveloped in a large cloud of hot gas (ICM), visible by its X-ray emission. One may assume that the electron density distribution associated with the X-ray brightness is in hydrostatic equilibrium, and one can extract the ICM radial density profiles by fits.

The amount of matter in the form of hot gas can be deduced from the intensity of this radiation. Adding the gas mass to the observed luminous matter, the total amount of baryonic matter, $M_b$, can be estimated \cite{Markevitch, Boni}.
In clusters studied, the gas fraction increases with the distance from
the center; the dark matter is more concentrated than the visible matter.

The temperature of the gas depends on the strength of the gravitational field, from which the total amount of gravitating matter, $M_{\rm grav}$, in the system can be deduced. In many such small galaxy groups one finds $M_{\rm grav}/M_b\geq3$. Thus a dark halo must be present. An accurate estimate of $M_{\rm grav}$ requires that also dark energy is taken into account, because it reduces the strength of the gravitational potential. There are sometimes doubts whether all galaxies apparently in these groups are physical members. If not, they will artificially increase the velocity scatter and thus lead to larger virial masses.

On the scale of large clusters of galaxies like the Coma, it is generally observed that dark matter represents about 85\% of the total mass and that the visible matter is mostly in the form of a hot ICM.
\section{Mass to luminosity ratios}

Let us define the mass/luminosity ratio of an astronomical object as $\Upsilon\equiv M/L$. Stellar populations exhibit values $\Upsilon=1-10$ in solar units, in the solar neighborhood $\Upsilon =2.5-7$, in the Galactic disk $\Upsilon =1.0-1.7$ \cite{Flynn}. In the Coma the value of $\Upsilon$ depends strongly on the halo profile for $r< 0.2$ Mpc; beyond that the value is 400 and slowly decreasing with radius \cite{Lokas}. Many dwarf spheroidal galaxies exhibit very high ratios: in Draco $\Upsilon=330\pm 125$, in Andromeda IX $\Upsilon= 93^{+120}_{-50}$. The cluster AC 114 discussed in Sec. 2.2 is very underluminous having $\Upsilon=700\pm 100$ \cite{Sereno}.

A special case is the ultra-faint dwarf disk galaxy Segue 1 \cite{Xiang} which has a baryon mass of only about 1000 solar masses. One interpretation is that it is a thin non-rotating stellar disk not accompanied by a gas disk, embedded in an axisymmetric DM halo and with an $f = M_{halo}/M_b \approx 200$. But if the disk rotates, $f$ could be as high as 2000. If Segue 1 also has a magnetized gas disk, the DM halo has to confine the effective pressure in the stellar disk and the magnetic Lorentz force in the gas disk as well as possible rotation. Then $f$ could be very large \cite{Xiang}.

These mass/luminosity ratios are however local, characteristic for isolated objects but not for the whole observable Universe. The value for $L_{\rm Universe}$ is roughly known. One can then deduce that the critical density to close the Universe would require $\Upsilon\approx 990$. However, there is not sufficiently much luminous matter to achieve closure, therefore a large amount of DM is needed.

   \begin{figure}[htbp]
   \includegraphics[width=13cm]{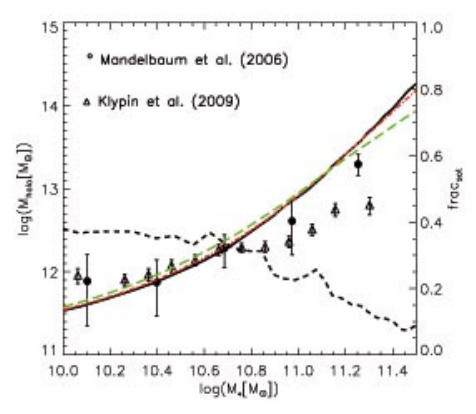}
   \caption{Dark matter halo mass $M_{halo}$ as a function of stellar mass $M_{\ast}$.
The thick black curve is the prediction from abundance matching
assuming no dispersion in the relation between the two masses. Red
and green dashed curves assume some dispersion in $\log M_{\ast}$. The dashed black curve is the satellite fraction as a function of stellar mass, as labeled on the axis at the right-hand side of the plot. From Qi Guo \& al. \cite{Guo}.}
      \end{figure}

\section{Mass autocorrelation functions}
If galaxy formation is a local process, then on large scales galaxies must trace mass. This requires the study of how galaxies populate dark matter haloes. In simulations one attempts to track galaxy and dark matter halo evolution across cosmic time in a physically consistent way, providing positions, velocities, star formation
histories and other physical properties for the galaxy populations of interest.

Guo \& {\em{al.}} \cite{Guo} use abundance matching arguments to derive an accurate relation between galaxy stellar mass and dark matter halo mass. They combine a
stellar mass function based on spectroscopic observations with a precise halo/subhalo mass function obtained from simulations. Assuming this stellar mass – halo mass relation to be unique and monotonic, they compare it with direct observational estimates of the mean mass of haloes surrounding galaxies of given stellar mass inferred from gravitational lensing and satellite galaxy dynamics data, and use it to populate haloes in simulations. The stellar mass – halo mass relation is shown in Fig. 5.
\begin{figure}[htbp]
   \includegraphics[width=15cm]{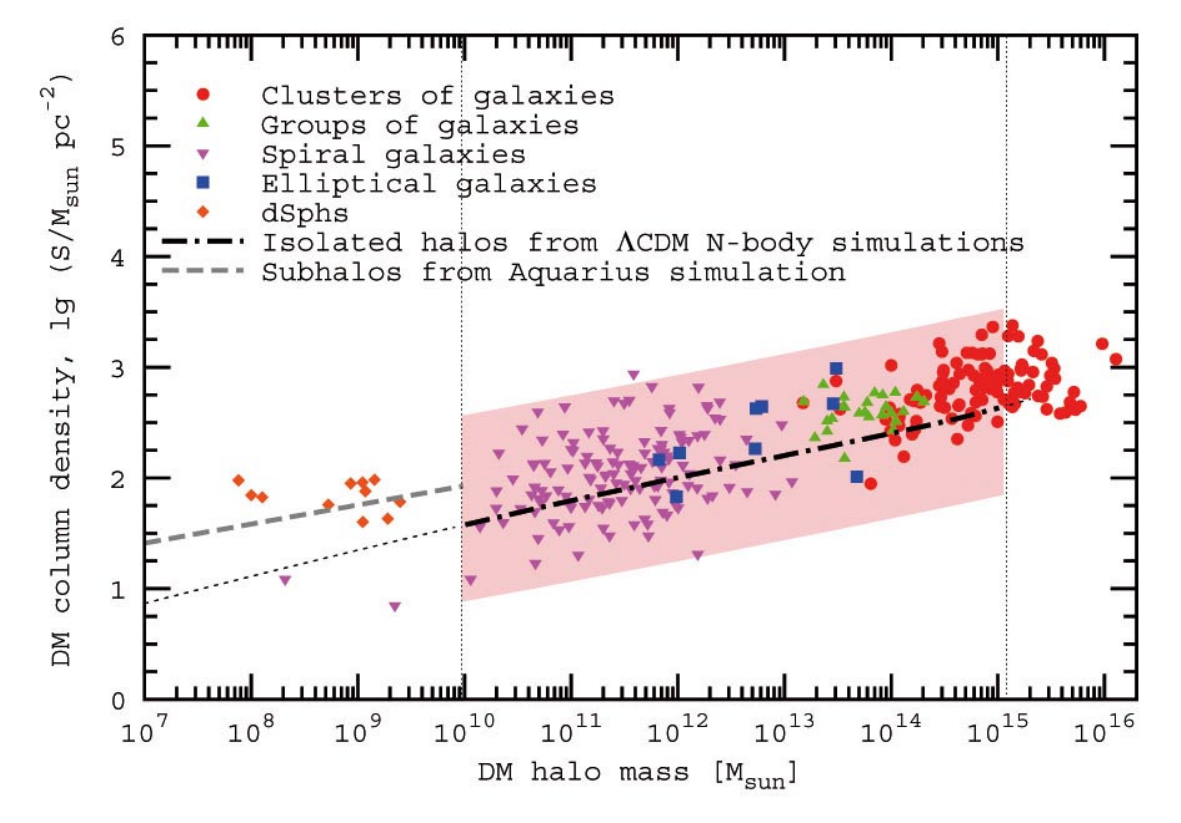}
   \caption{Dark matter column density vs. dark matter halo mass in solar units.  From A. Boyarsky \& {\em al.} \cite{Boyarsky}}
      \end{figure}
The implied spatial clustering of stellar
mass turns out to be in remarkably good agreement with a direct and
precise measurement. By comparing the galaxy autocorrelation function with the total mass autocorrelation function, as averaged over the Local Supercluster (LSC) volume, one concludes that a large amount of matter in the LSC is dark.

A similar study is that of Boyarsky \& al. \cite{Boyarsky} who find a universal relation between DM column density and DM halo mass, satisfied by matter distributions at all observable scales, in halo sizes from $10^8$ to $10^{16}$ solar masses, as shown in Fig. 6. Such a universal property is difficult to explain without dark matter.

\section{Strong and weak lensing}
A consequence of the Strong Equivalence Principle (SEP) is that a photon in a
gravitational field moves as if it possessed mass, and light rays therefore bend around gravitating masses. Thus celestial bodies can serve as \textit{gravitational lenses} probing the gravitational field, whether baryonic or dark without distinction.

Since photons are neither emitted nor absorbed in the process of gravitational light deflection, the surface brightness of lensed sources remains unchanged. Changing the size of the cross-section of a light bundle only changes the flux observed from a source and magnifies it at fixed surface-brightness level. If the mass of the lensing object is very small, one will merely observe a magnification of the brightness of the
lensed object an effect called \textit{microlensing}. Microlensing of distant quasars by compact lensing objects (stars, planets) has also been observed and used for estimating the mass distribution of the lens--quasar systems.
   \begin{figure}[htbp]
   \includegraphics[width=13cm]{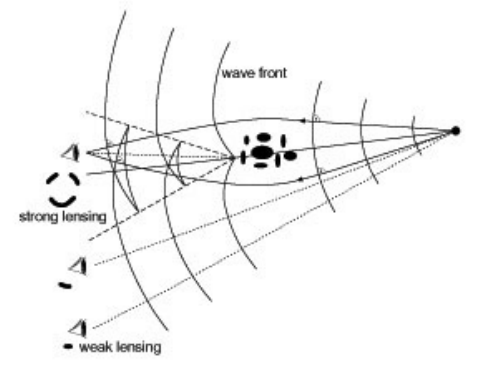}
   \caption{Wave fronts and light rays in the presence of a cluster perturbation. From N. Straumann \cite{Straumann}.}
      \end{figure}

\textit{Weak Lensing} refers to deflection through a small angle when the light ray can be treated as a straight line (Fig. 7), and the deflection as if it occurred discontinuously at the point of closest approach (the thin-lens approximation in optics). One then only invokes SEP which accounts for the distortion of clock rates. \begin{figure}[htbp]
   \includegraphics[width=9cm]{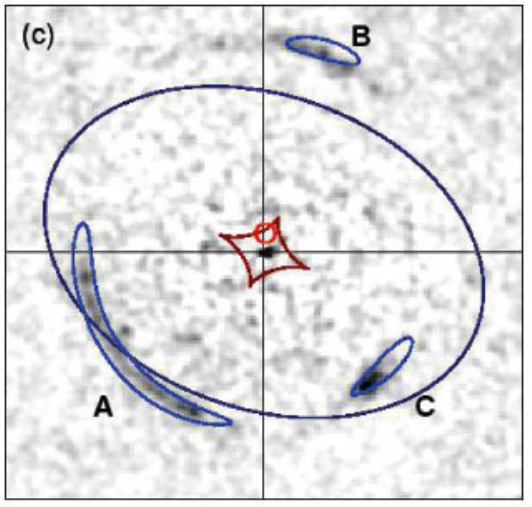}
   \caption{This image resulted from color-subtracteion of a lensing singular isothermal elliptical galaxy. The strongly lensed object forms two prominent arcs A, B and a less extended third image C. From R.J. Smith \& {\em al.} \cite{Smith}}
      \end{figure}
 \begin{figure}[htbp]
   \includegraphics[width=11cm]{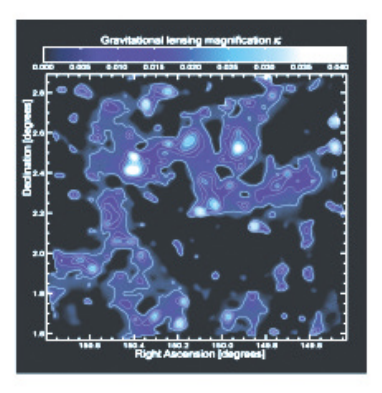}
   \caption{Map of the dark matter distribution in the 2-square degree
COSMOS field \cite{Massey}: the linear blue scale on top shows the gravitational lensing magnification $\kappa$, which is proportional to the projected mass along the line of sight. Contours begin at 0.4\% and are spaced by 0.5\% in $\kappa$.}
      \end{figure}

In \textit{Strong
Lensing} the photons move along geodesics in a strong gravitational potential which distorts space as well as time, causing larger deflection angles and requiring the full theory of GR. The images in the observer plane can then become quite complicated because there may be more than one null geodesic connecting source and observer; it may not even be possible to find a unique mapping onto the source plane {\em c.f.} Fig.7 . Strong lensing is a tool for testing the distribution of mass in the lens rather than purely a tool for testing GR. An illustration is seen in Fig. 8 where the lens is an elliptical galaxy \cite{Smith}.

At cosmological distances one may observe lensing by composed objects such as galaxy groups which are ensembles of ``point-like'', individual galaxies. Lensing effects are very model-dependent, so to learn the true magnification effect one
needs very detailed information on the structure of the lens.
\begin{figure}[htbp]
   \includegraphics[width=12cm]{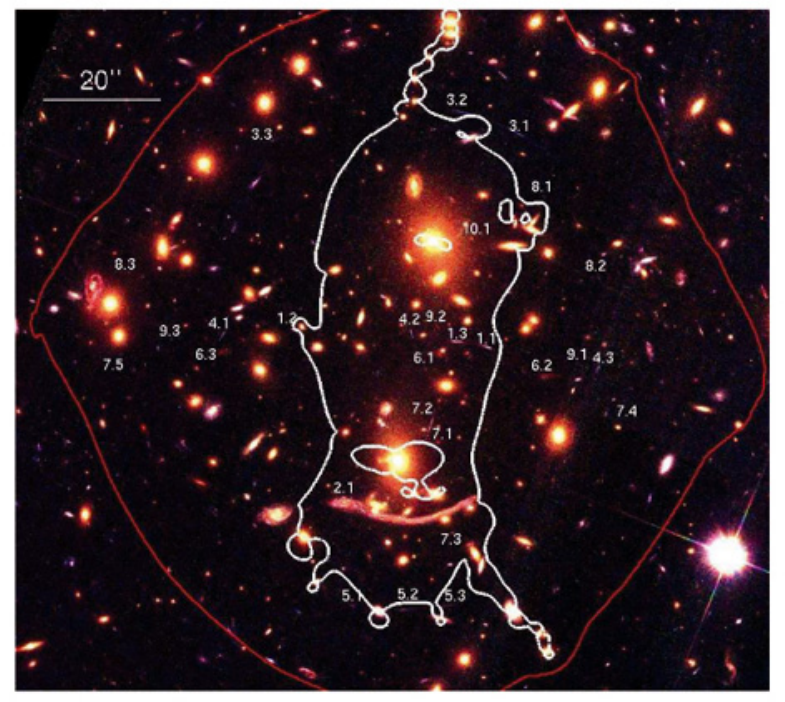}
   \caption{Strong lensing analysis of the galaxy cluster Abell 370 (z=0.375) shown with the locations of the identified multiple systems within the critical line
at z=1.2 in white (the redshift for the majority of the mutiple images). The red line delimits the region of multiple images for very high redshift
sources (here assuming z = 6). From J. Richard \& {\em al.} \cite{Richard}}
      \end{figure}

The large-scale distribution of matter in the Universe is inhomogeneous in every direction, so one can expect that everything we observe is displaced and distorted by weak lensing. Since the tidal gravitational field and the
deflection angles depend neither on the nature of the matter nor on its physical state, light deflection probes the total projected mass distribution. Lensing in infrared light offers an additional advantage of being able to sense distant background galaxies, since their number density is higher than in the optical range.

Background galaxies would be ideal tracers of distortions if they were intrinsically circular, because lensing transforms circular sources into ellipses. Any measured ellipticity would then directly reflect the action of the gravitational tidal field of the interposed lensing matter, and the statistical properties of the distortions would reflect the properties of the matter distribution. But many galaxies are actually intrinsically elliptical, and the ellipses are randomly oriented. This introduces noise into the inference of the tidal field from observed ellipticities. A useful feature in the sky is a fine-grained pattern of faint and distant blue galaxies appearing as a `wall paper'. This makes statistical weak-lensing studies possible, because it allows the detection of the coherent distortions imprinted by gravitational lensing on the images of the galaxy population. Thus weak lensing has become an important technique to map non-luminous matter. A reconstruction of the largest and most detailed weak lensing survey ever undertaken with the Hubble Space Telescope is shown in Fig. 9 \cite{Massey}. This map covers a large enough area to see extended filamentary structures.

In Fig. 10 we see a first detailed example of the strong lensing of a cluster \cite{Richard}.

\section{Cosmic Microwave Background (CMB)}
\voffset= 0.4cm

\begin{figure}[htbp]
   \includegraphics[width=14.2cm]{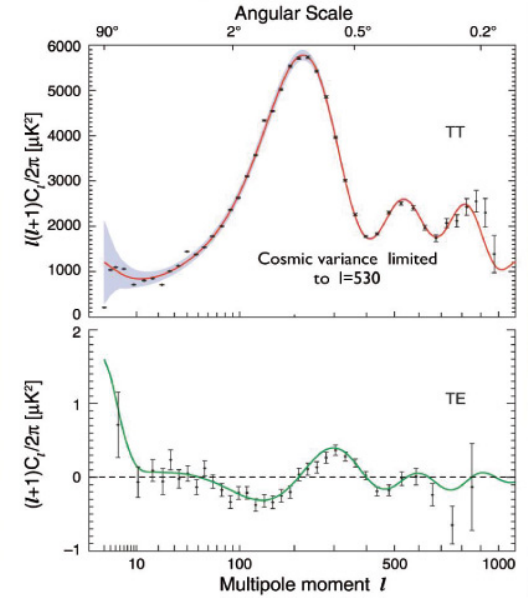}
   \caption{The maps of CMB radiation temperature (TT) and temperature-polarization correlations (TE) from WMAP show anisotropies which can be analyzed by power spectra as functions of multipole moments.  From G. Hinshaw \& {\em al.} \cite{Hinshaw}}
      \end{figure}

The tight coupling between radiation and matter density before decoupling
caused the primordial adiabatic perturbations to oscillate in phase.
Beginning from the time of last scattering, the receding horizon has been revealing these
frozen density perturbations, setting up a pattern of standing
acoustic waves in the baryon--photon fluid.  After
decoupling, this pattern is visible today as temperature anisotropies with a certain regularity across the sky.

The primordial photons are polarized by the anisotropic Thomson
scattering process, but as long as the photons continue to meet free
electrons their polarization is washed out, and no net polarization
is produced. At a photon's last scattering, however, the induced
polarization remains and the subsequently free-streaming photon
possesses a quadrupole moment.

Temperature and polarization fluctuations are analyzed in terms of multipole components or powers. The resulting distribution of powers versus multipole $\ell$, or multipole moment $k=2\pi/\ell$, is the \textit{power spectrum} which exhibits conspicuous \emph{Doppler peaks}. In Fig. 11 we display the TT and TE power spectra from the five-year data of WMAP as functions of multipole moments \cite{Hinshaw}. The spectra can then be compared to theory, and theoretical parameters determined. Many experiments have determined the power spectra, so a wealth of data exists.
\begin{figure}[htbp]
   \includegraphics[width=12cm]{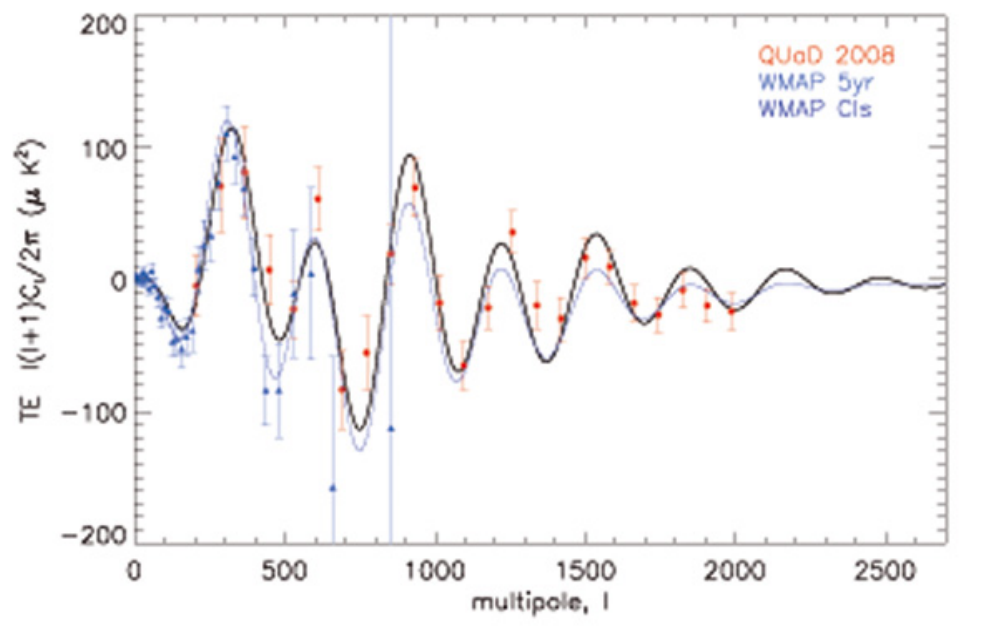}
   \caption{The E-mode polarization power spectrum (EE) from the CMB observations of the QUaD collaboration \& {\em al.} \cite{QUAD}}
      \end{figure}

Baryonic matter feels attractive self-gravity and is pressure-supported, whereas dark matter only feels attractive self-gravity, but is pressureless. Thus the Doppler peaks in the CMBR power spectrum testify about baryonic and dark matter, whereas the troughs testify about rarefaction caused by the baryonic pressure. The position of the first peak determines $\Omega_m h^2$. Combining the 5-year WMAP measurements of the temperature power spectrum (TT) with determinations of the Hubble constant $h$, the WMAP team finds the total mass density parameter $\Omega_m\approx 0.26$ \cite{Komatsu}. The ratio of amplitudes of the second-to-first Doppler peaks determines the baryonic density parameter $\Omega_b\approx 0.04$; the dark matter component is then $\Omega_{\rm dm}\approx 0.22$.

Information on the temperature -- E-mode polarization correlations (TE) comes from several measurements, among them WMAP \cite{Hinshaw, Komatsu}, and on the E-mode polarization power spectrum alone (EE) from the QUAD collaboration \cite{QUAD}, Fig. 12.

The results show two surprises: Firstly, since $\Omega_m\ll1$, a large component $\Omega_{\Lambda}\approx0.74$ is missing,
of unknown nature, and termed \textit{dark energy}. The second
surprise is that ordinary baryonic matter is only a small fraction
of the total matter budget. The remainder is then dark matter, of unknown composition.
Of the 4.2\% of baryons in the Universe only about 1\% is stars.

\section{Baryonic acoustic oscillations (BAO)}

A cornerstone of cosmology is the Copernican principle, that matter in the Universe is distributed homogeneously, if only on the largest scales of superclusters separated by voids. On smaller scales we observe inhomogeneities in the forms of galaxies, galaxy groups, and clusters. The common approach to this situation is to turn to non-relativistic hydrodynamics and treat matter in the Universe as an adiabatic, viscous, non-static fluid, in which random fluctuations around the mean density appear, manifested by compressions in some regions and rarefactions in other. The origin of these density fluctuations was the tight coupling
established before decoupling between radiation and charged matter density, causing them to oscillate in phase. An ordinary fluid is dominated by the material pressure, but in the fluid of our Universe three effects are competing: gravitational attraction, density dilution due to the Hubble flow, and radiation pressure felt by charged particles only.

The inflationary fluctuations crossed the post-inflationary Hubble radius, to come back into vision with a wavelength corresponding to the size of
the Hubble radius at that moment. At time $t_{eq}$ the overdensities began to amplify and grow into larger inhomogeneities. In overdense regions where the gravitational forces dominate, matter contracts locally and attracts surrounding matter, becoming increasingly unstable until it eventually collapses into a gravitationally bound object. In regions where the pressure forces dominate, the fluctuations move with constant amplitude as sound waves in the fluid, transporting energy from one region of space to another.

Inflationary models predict that the primordial mass density fluctuations should be adiabatic, Gaussian, and exhibit the same scale invariance as the CMB fluctuations.
The baryonic acoustic oscillations can be treated similarly to CMB, they are specified by the dimensionless \textit{mass autocorrelation function} which is the Fourier transform of the power spectrum of a spherical harmonic expansion. The power spectrum is shown in Fig. 13 \cite{BAO}.

As the Universe approached decoupling, the photon mean free path increases
and radiation can diffused from overdense regions into underdense ones, thereby smoothing out any inhomogeneities in the plasma. The situation changed dramatically at recombination, at time 380\,000~yr after Big Bang, when all the free electrons suddenly disappeared, captured into atomic Bohr orbits, and the radiation pressure almost vanished. Now the baryon acoustic waves and the CMB continued to oscillate independently, but adiabatically, and the density perturbations which have entered the Hubble radius since then can grow with full vigor into baryonic structures.

The scale of BAO depends on $\Omega_m$ and on the Hubble constant, $h$, so one needs information on $h$ to break the degeneracy. The result is then $\Omega_m\approx0.26$. In the ratio $\Omega_b/\Omega_m$ the $h$-dependence cancels out, so one can also quantify the amount of dark matter on very large scales by $\Omega_b/\Omega_m=0.18\pm 0.04$.
\begin{figure}[htbp]
   \includegraphics[width=12cm]{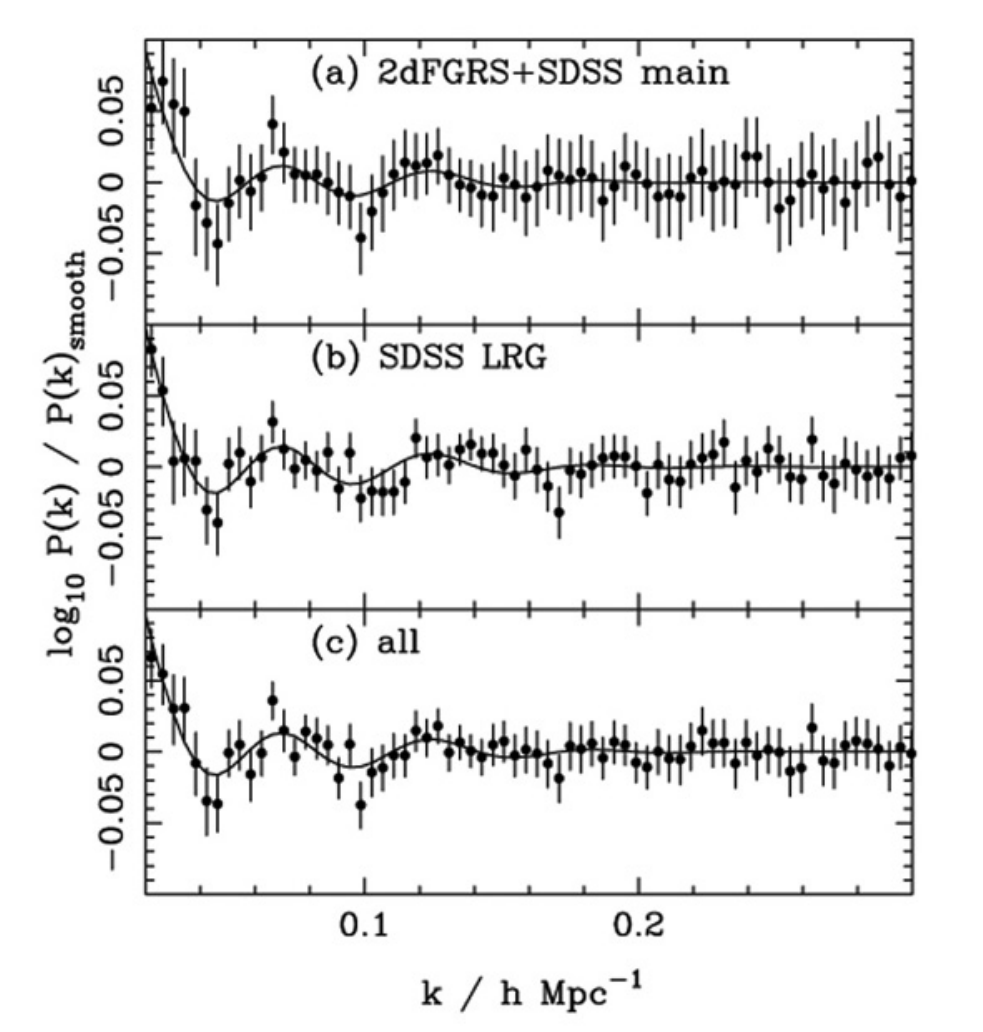}
   \caption{BAO in power spectra calculated from (a) the combined SDSS
and 2dFGRS main galaxies, (b) the SDSS DR5 LRG sample, and (c) the
combination of these two samples (solid symbols with $1\sigma$ errors). The data
are correlated and the errors are calculated from the diagonal terms in the
covariance matrix. A Standard $\Lambda$CDM distance-redshift relation was assumed
to calculate the power spectra with $\Omega_m = 0.25,~\Omega_{\Lambda} = 0.75$. From Percival \& {\em al.} \cite{BAO}.}
      \end{figure}

\section{Galaxy formation in purely baryonic matter}
We have seen in Sec. 9 that the baryonic density parameter, $\Omega_b$, is very small. The critical density $\Omega_{critical}$ is determined by the expansion speed of the Universe, and the mean baryonic density of the Universe (stars, interstellar \& intergalactic gas) is only $\Omega_b = 0.046\times\Omega_{critical}$ \cite{Komatsu}.

The question arises whether the galaxies could have formed from primordial density fluctuations in a purely baryonic medium. We have also noted, that the fluctuations in CMB and BAO maintain adiabaticity. The amplitude of the primordial baryon density fluctuations would have needed to be very large in order to form the observed number of galaxies. But then the amplitude of the CMB fluctuations would also have been very large, leading to intolerably large CMB anisotropies today. Thus galaxy formation in purely baryonic matter is ruled out by this argument alone.

Thus one concludes, that the galaxies could only have been formed in the presence of gravitating dark matter which started to fluctuate early, unhindered by radiation pressure. This conclusion is further strengthened in the next Section.

\section{Large Scale Structures simulated}
In the $\Lambda$CDM paradigm, the
nonlinear growth of dark matter structure is a well-posed
problem where both the initial conditions and the evolution
equations are known (at least when the effects of the baryons can be neglected).

The Aquarius Project \cite{Aquarius} is a Virgo Consortium programme to carry out
high-resolution dark matter simulations of Milky-Way-sized
halos in the $\Lambda$CDM cosmology. This project seeks clues to
the formation of galaxies and to the nature of the dark matter
by designing strategies for exploring the formation of our
Galaxy and its luminous and dark satellites.
\begin{figure}[htbp]
   \includegraphics[width=15cm]{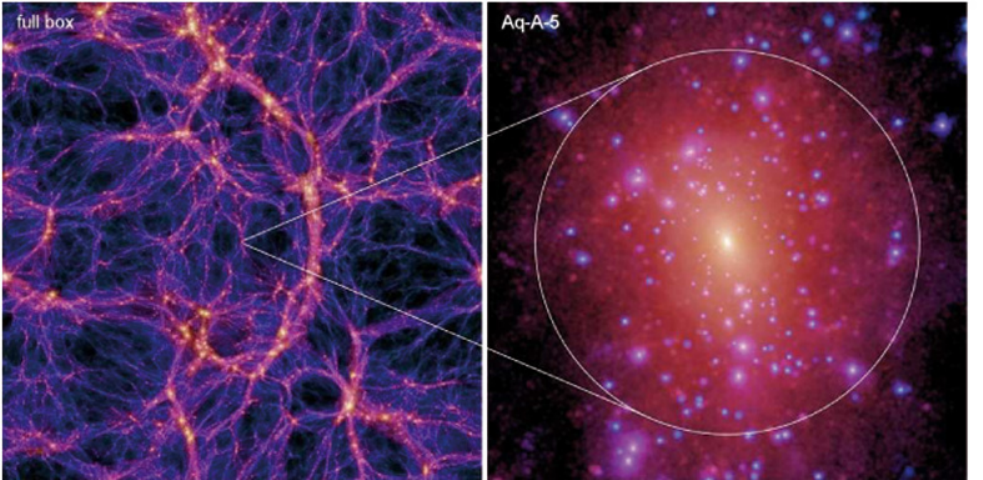}
   \caption{The left panel shows the projected dark matter density at z = 0 in a slice of thickness 13.7 Mpc through the full box
(137 Mpc on a side) of the $900^3$ parent simulation. The right panel
show this halo resimulated at a different numerical resolution. The image brightness is proportional to the logarithm of the squared dark matter density
projected along the line-of-sight. The circles mark $r_{50}$, the radius within which the mean density is 50 times the background density. From V. Springel \& al. \cite{Aquarius}}
\end{figure}
The galaxy population on scales from 50 kpc to the size of the observable Universe has been predicted by hierarchical $\Lambda$CDM scenarios, and compared directly with a wide array of observations. So far, the $\Lambda$CDM paradigm has passed these tests successfully, particularly those that consider the large-scale matter distribution and has led to the discovery of a universal internal structure for dark matter halos
As was noted already in Sec. 11, the observed structure of galaxies, clusters and superclusters, as illustrated by Fig. 14,  could not have formed in a baryonic medium devoid of dark matter.

Given this success, it is important to test $\Lambda$CDM predictions
also on smaller scales, not least because these are sensitive
to the nature of the dark matter. Indeed, a number of serious
challenges to the paradigm have emerged on the scale
of individual galaxies and their central structure. In particular,
the abundance of small dark matter subhalos predicted
within CDM halos is much larger than the number of known satellite galaxies
surrounding the Milky Way.

\section{Dark matter from overall fits}
In Sec. 9 we have seen that the WMAP-5 year CMB data together with the Hubble constant value testify about the existence of dark matter \cite{Hinshaw, Komatsu}. In Sec. 10 we addressed the BAO data \cite{BAO} with the same conclusion. In overall fits one combines these with supernova data (SN Ia) which offer a constraint nearly orthogonal to that of CMB in the $\Omega_{\Lambda}-\Omega_m$-plane.
\begin{figure}[htbp]
   \includegraphics[width=13cm]{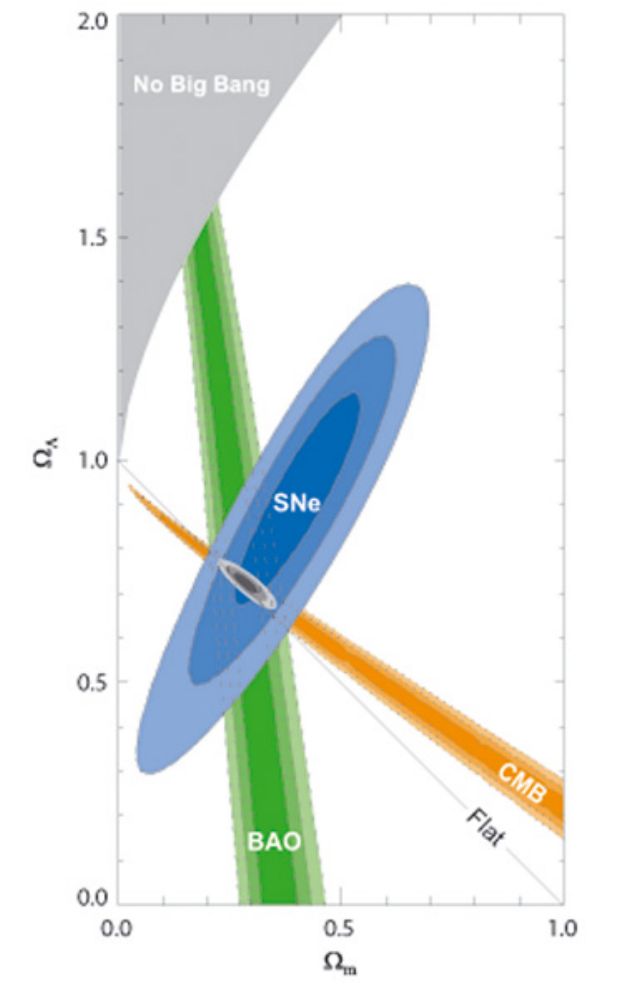}
   \caption{68.3 \%, 95.4 \% and 99.7\% confidence level contours on $\Omega_{\Lambda}$ and $\Omega_m$ obtained from CMB, BAO and the Union SN set, as well as their combination (assuming w = -1). Note the straight line corresponding to a flat Universe with $\Omega_{\Lambda}+\Omega_m=1$. From M. Kowalski \& al. \cite{Kowalski}.}
\end{figure}
The "Union" compilation of 414 SN Ia, which reduces to 307 SNe after selection cuts, includes the recent large samples of SNe Ia from the Supernova Legacy Survey and ESSENCE Survey, the older data sets, as well as the recently extended data set of distant supernovae observed with HST. M. Kowalski \& al. \cite{Kowalski} present the latest results from this Union compilation and discuss the cosmological constraints and its combination with CMB and BAO measurements. The CMB constraint is close to the line $\Omega_{\Lambda}+\Omega_m=1$, whereas the supernova constraint is close to the line $\Omega_{\Lambda}-1.6\times\Omega_m=0.2$. The BAO data constrain $\Omega_m$, but hardly at all $\Omega_{\Lambda}$. This is shown in Fig. 15.

Defining the vacuum energy density parameter by
\begin{equation}
\Omega_k=1-\Omega_{\Lambda}-\Omega_m~,
\end{equation}
a flat Universe corresponds to $\Omega_k=0$. For a $\Lambda CDM$ Universe with a cosmological constant responsible for dark energy, a simultaneous fit to the data sets gives
\begin{equation}
\Omega_m=0.285^{+0.020}_{-0.019}\pm 0.011,~~~~
\Omega_k=-0.009^{+0.009+0.002}_{-0.010-0.003}\ ,
\end{equation}
where the first error is statistical and the second error systematic. This is the most precise determination at present of the total mass density parameter $\Omega_m$. At the same time one notes that the Universe is consistent with being flat. Subtracting $\Omega_b = 0.046$ from $\Omega_m=0.285$ one obtains the density parameter for dark matter, $\Omega_{dm}\approx 0.24$. Assuming flatness, M. Kowalski \& al.  \cite{Kowalski} find $\Omega_m=0.274\pm 0.016\pm 0.013$.

\begin{figure}[htbp]
   \includegraphics[width=13cm]{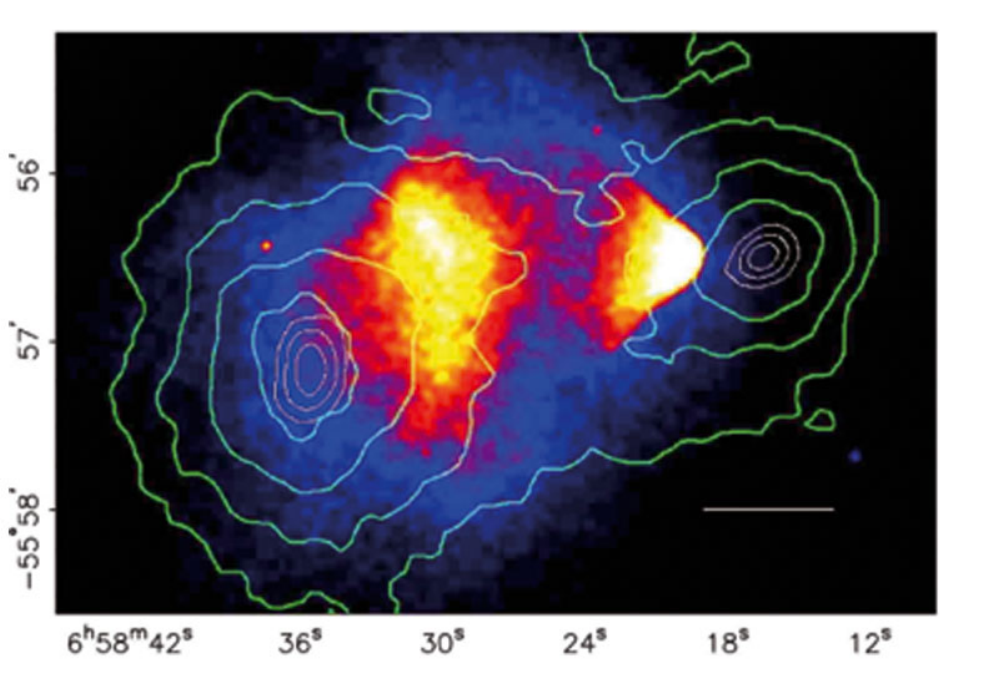}
   \caption{The merging cluster 1E0657-558. On the right is the smaller 'bullet' cluster which has traversed the larger cluster. The colors indicate the X-ray temperature of the plasma: blue is coolest and white is hottest. The green contours are the weak lensing reconstruction of the gravitational potential of the cluster. From D. Clowe \& al. \cite{Clowe}}
\end{figure}

\section{Merging galaxy clusters}
A direct empirical proof of the existence of dark matter is furnished by observations of 1E0657-558, a unique cluster merger \cite{Clowe}. Due to the collision of two clusters, the dissipationless stellar component and the fluid-like X-ray emitting plasma are spatially segregated. The gravitational potential observed by weak and strong lensing does not trace the plasma distribution which is the dominant baryonic mass component, but rather approximately traces the distribution of galaxies, cf. Fig. 16. The center of the total mass is offset from the center of the baryonic mass peaks, proving that the majority of the matter in the system is unseen. In front of the smaller 'bullet' cluster which has traversed the larger one, a bow shock is evident in the X-rays.

Two other merging systems with similar characteristics have been seen, although with lower spatial resolution and less clear-cut cluster geometry. In Fig.17 we show the post-merging galaxy cluster MACS J0025.4-1222 \cite{Bradac}.

\begin{figure}[htbp]
   \includegraphics[width=13cm]{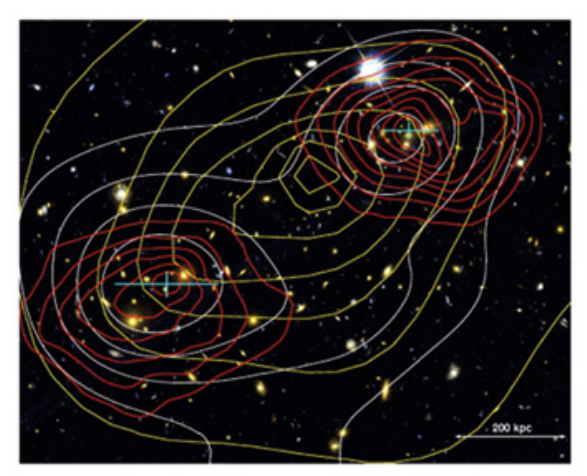}
   \caption{The F555W-F814W color composite of the cluster MACS J0025.4.1222. Overlaid in red contours is the surface mass density $\kappa$ from the
combined weak and strong lensing mass reconstruction. The contour levels are linearly spaced with $\Delta\kappa = 0.1$, starting at  $\Delta\kappa= 0.5$, for a
fiducial source at a redshift of $z_s\rightarrow\infty$. The X-ray brightness contours (also linearly spaced) are overlaid in yellow and the I-band light is
overlaid in white. The measured peak position and error bars for the total mass are given as a cyan cross \cite{Bradac}.}
\end{figure}

\section{Comments and conclusions}
What we have termed ``dark matter'' is generic for observed gravitational effects
on all scales: galaxies, small and large galaxy groups, clusters and superclusters, CMB anisotropies over the full horizon, baryonic oscillations over large scales, and cosmic shear in the large-scale matter distribution. The correct explanation or nature of dark matter is not known, whether it implies conventional matter, unconventional particles, or modifications to gravitational theory. but gravitational effects prove its existence in some form.

Only a few per cent of the total mass of the Universe is found in stars and hydrogen clouds, and this amount of baryonic matter is well accounted for by nucleosynthesis. If there exist particles which were very slow at time $t_{\rm eq}$ when galaxy formation started, they could be candidates for cold dark matter. If dark matter were composed of such new particles, they must have become non-relativistic much earlier than the leptons, and then decoupled from the hot plasma.

Whenever laboratory searches discover a new particle, it must pass several tests in order to be considered a viable DM candidate: it must be neutral, compatible with constraints on self-interactions (essentially collisionless), consistent with Big Bang nucleosynthesis, and match the appropriate relic density. It must be consistent with direct DM searches and gamma-ray constraints, it must leave stellar evolution unchanged, and be compatible with other astrophysical bounds.

Theories attempting to explain dark matter without new particles, with new interactions or modified gravity, likewise have the burden to explain all the observed gravitational effects described in here. Thus there remains much to be done.

\section*{Acknowledgements}
It is a pleasure to thank K. Mattila at the University of Helsinki, M. Valtonen at the University of Turku, and P. Salucci at SISSA, Trieste, for valuable comments.


\section*{References}\begin{thebibliography}{99}
\bibitem{Oort1} J. H. Oort, {\em The force exerted by the stellar system in the direction perpendicular to the galactic plane and some related problems}, Bull. Astron. Inst. Netherlands, {\bf 6}, 249 (1932)
\bibitem{Zwicky} F. Zwicky, {\em Die Rotverschiebung von extragalaktischen Nebeln}, Helvetica Physica Acta, {\bf 6}, 110 (1933)
\bibitem{Holmberg} J. Holmberg \& C. Flynn, {\em The local density of matter mapped by Hipparchos}, \mnras, {\bf 313}, 209 (2000)
\bibitem{Oort2} J. H. Oort, {\em Infall of Gas from Intergalactic Space}, Nature {\bf 224}, 1158 (1969)
\bibitem{NFW} J. F. Navarro, C. S. Frenk \& S. D. M. White, {\em A Universal Density Profile from Hierarchical Clustering}, Astrophys. J. {\bf 490}, 493 (1997)
\bibitem{Moore} B. Moore \& {\em al., Resolving the Structure of Cold Dark Matter Haloes}, Astrophys. J., {\bf 499}, L5 (1998)
\bibitem{Gentile} G. Gentile\& {\em al., The cored distribution of dark matter in spiral galaxies}, \mnras {\bf 351}, 903 (2004)
\bibitem{Lokas} E. L. Lokas \& G. Mamon, {\em Dark matter distribution in the Coma cluster from galaxy kinematics: breaking the mass-anisotropy degeneracy}, \mnras {\bf 343}, 401 (2003)
\bibitem{Cupani} G. Cupani, M. Mezzetti \& F. Mardirossian, {\em Cluster mass estimation through fair galaxies}, arXiv:0910.2882[astro-ph.CO] (2009)
\bibitem{Coma} M. J. Valtonen \& G. G. Byrd, {\em A binary model for the Come cluster of galaxies}, Astrophys. J., {\bf 230}, 655 (1979)
\bibitem{Bradac} M. Brada$\check{c}$ \& {\em al., Revealing the properties of dark matter in the merging cluster MACSJ0025.4-1222}, Astrophys. J., {\bf 687}, 959 (2008)
\bibitem{Allen} S. W. Allen \& {\em al., Improved constraints on dark energy from Chandra X-ray observations of the largest relaxed galaxy clusters}, \mnras, {\bf 383}, 879 (2008)
\bibitem{Limousin} M. Limousin \& {\em al., Strong Lensing in Abell 1703: Constraints on the Slope of the Inner Dark Matter Distribution}, A\& A, {\bf 489}, 23 (2008)
\bibitem{Sereno} M. Sereno, M. Lubini \& Ph. Jetzer, {\em A multi-wavelength strong lensing analysis of baryons and dark matter in the dynamically active cluster AC 114}, arXiv:0904.0018[astro-ph.CO] (2009)
\bibitem{Valtonen} A. D. Chernin \& {\em al., Dark energy and the mass of the Local Group}, arXiv:0902.3871[astro-ph.CO] (2009)
\bibitem{Salucci} P. Salucci \& al., {\em The Universal Rotation Curve of Spiral Galaxies. II The Dark Matter Distribution out to the Virial Radius}, \mnras, {\bf 378}, 41 (2007), and P. Salucci, {\em The mass distribution in Spiral galaxies}, arXiv:0707.4370[astro-ph] (2007)
\bibitem{Sofue} Y. Sofue, M. Honma \& T. Omodaka, {\em Unified Rotation Curve of the Galaxy — Decomposition into de Vaucouleurs Bulge, Disk, Dark Halo, and the 9-kpc Rotation Dip}, Publ. Astron. Soc. Japan, {\bf 61}, 227 (2009) and arXiv:0811.0859[astro-ph] (2008)
\bibitem{Ullio} R. Catena \& P. Ullio, {\em A novel determination of the local dark
    matter density}, arXiv:0907.0018[astro-ph.CO] (2009)
\bibitem{Weber} M. Weber \& W. de Boer, {\em Determination of the Local Dark Matter Density in our Galaxy}, arXiv:0910.4272[astro-ph.CO] (2009)
\bibitem{Markevitch} M. Markevitch \& {\em al., Mass Profiles of the Typical Relaxed Galaxy Clusters A2199 and A496}, Astrophys. J., {\bf 527}, 545 (1999)
\bibitem{Boni} C. De Boni \& G. Bertin, {\em The relative concentration of visible and dark matter in clusters of galaxies}, arXiv:0805.0494[astro-ph] (2008)
\bibitem{Flynn} C. Flynn \& {\em al., On the mass-to-light ratio of the local Galactic disk and the optical luminosity of the Galaxy}, \mnras, {\bf 372}, 1149 (2006)
\bibitem{Xiang} M. Xiang-Gruess, Y.-Q. Lou \& W. J. Duschl, {\em Dark matter dominated dwarf disc galaxy Segue 1}, arXiv:0909.3496[astro-ph.CO] (2009)
\bibitem{Guo} Qi Guo \& {\em al., How do galaxies populate dark matter haloes?}, arXiv:0909.4305[astro-ph.CO] (2009)
\bibitem{Boyarsky} A. Boyarsky \& {\em al., New evidence for dark matter}, arXiv:0911.1774[astro-ph.CO] (2009)
\bibitem{Straumann} N. Straumann, {\em Matter in the Universe}, Space Science Series of ISSI, vol. 14. Kluwer (2002) (Reprinted from Space Sci. Rev., {\bf{100}}, 29(2002).)
\bibitem{Smith} R.J. Smith \& {\em al., Discovery of Strong Lensing by an Elliptical Galaxy at z=0.0345}, Astrophys.J., {\bf 625}, L103 (2005)
\bibitem{Massey} R. Massey \& {\em al, Dark matter maps reveal cosmic scaffolding}, Nature, {\bf 445}, 286 (2007)
\bibitem{Richard} J. Richard \& {\em al, Abell 370 revisited: refurbished Hubble imaging of the first strong lensing cluster}, arXiv:0910.5553[astro-ph.CO] (2009)
\bibitem{Hinshaw} G. Hinshaw \& {\em al, Five-Year Wilkinson Microwave Anisotropy Probe (WMAP) Observations: Data Processing, Sky Maps, and Basic Results }, Astrophys.J.Suppl. {\bf 180}, 225 (2009)
\bibitem{Komatsu} E. Komatsu  \& {\em al, Five-Year Wilkinson Microwave Anisotropy Probe (WMAP) Observations: Cosmological interpretation}, Astrophys.J.Suppl. {\bf 180}, 330 (2009)
\bibitem{QUAD} QUad Collaboration -- P. G. Castro \& {\em al, Cosmological parameters from the QUAD CMB polarization experiment}, arXiv:0901.0810
    [astro-ph.CO] (2009)
\bibitem{BAO} W. J. Percival \& {\em al, Measuring the Baryon Acoustic Oscillation scale using the SDSS and 2dFGRS}, \mnras, {\bf 381}, 1053 (2007)
\bibitem{Aquarius} V. Springel  {\em al., The Aquarius Project: the subhalos of galactic halos}, \mnras, {\bf 391}, 1685 (2008)
\bibitem{Kowalski} M. Kowalski \& {\em al, Improved cosmological constraints from new, old and combined supernova datasets}, arXiv:0804.4142 [astro-ph] (2008)
\bibitem{Clowe} D. Clowe \& {\em al., A direct empirical proof of the existence of dark matter}, Astrophys. J. {\bf 648}, L109 (2006)

\end{thebibliography}
\end{document}